\documentclass[10pt]{amsart}
\usepackage{amsmath}
\usepackage{graphicx}
\usepackage{xspace}
\numberwithin{equation}{section}
\newcommand{\vsf}{$\mathcal{K}$ flow\xspace}
\renewcommand{\vec}[1]{\boldsymbol{#1}}
\renewcommand{\exp}[1]{e^{#1}}
\newcommand{\ud}[1]{\mathrm{d}#1}
\newcommand{\br}[1]{\left(#1\right)}
\newcommand{\sq}[1]{\left[#1\right]}

\newcommand{\upd}[3]{{#1}^{#2}_{\phantom{#2}{#3}}}

\newcommand{\om}{\tilde{\omega}}

\begin{document}
\begin{quote}
\begin{center}
{\textsc{\huge A class of space-times}}\\
\vspace{2ex}\textsc{\huge of non-rigidly rotating dust}%\\
\vspace{6ex}
\begin{quote}
\begin{center}{\large \textsc{{\L}ukasz Bratek}} \\
\emph{\small Institute of Nuclear Physics PAN, Radzikowskego 152,
31-342 Krak{\'o}w, Poland. } \emph{\small e-mail}: \texttt{\small
lukasz.bratek@ifj.edu.pl}\\
\vspace{1ex}
{\large \textsc{Joanna Ja{\l}ocha}} \\
\emph{\small Jagellonian University, Institute of Physics,
Reymonta 4, 30-059 Krak{\'o}w, Poland.} \emph{\small
e-mail}: \texttt{\small }\\
 \vspace{1ex}
{\large \textsc{Marek Kutschera}} \\
\emph{\small Institute of Nuclear Physics PAN, Radzikowskego 152,
31-342 Krak{\'o}w, Poland, \& Jagellonian University, Institute of
Physics, Reymonta 4, 30-059 Krak{\'o}w, Poland} \emph{\small
e-mail}: \texttt{\small marek.kutschera@ifj.edu.pl}
\end{center}
\end{quote}
\end{center}
\vspace{4ex}
%\begin{quote}
\begin{scriptsize}
{\bf Abstract. We find a class of exact solutions of
differentially rotating dust in the framework of General
Relativity. There exist asymptotically flat space-times of the
flow with positive mass function that for radii sufficiently large
is monotonic and tends to zero at infinity. Some of the
space-times may have non-vanishing total angular momentum. The
flow is essentially different from another exactly solvable flow
described by van Stockum line element.}
\end{scriptsize}
\end{quote}
\vspace{5ex}
\section{Introduction}
The paper is devoted to investigation of solutions of
self-gravitating cylindrically symmetric dust flow on circular
orbits along field lines of locally non-rotating observers in the
framework of General Relativity. The flow is non-rigid (differential) and non-expanding. It turns out the resulting
space-times are globally regular, apart from the points where
singularities may be located, and that the congruence of locally
non-rotating observers used to define energy-momentum exist
globally, as well. Consequently, the problem of the flow is well
posed. We shall call such a flow the {\bf \vsf} for brevity. The
main drawback of the flow is that the proper energy density is
negative definite and that the flow is purely relativistic -- has
no newtonian limit.

The result of our paper is the construction of the multipole
sequence of asymptotically flat, stationary, and cylindrically
symmetric external solutions to which a broad class of other
asymptotically flat solutions can be decomposed. We construct also
the corresponding multipole sequence of internal space-times. The
multipole representation is enumerable by a discrete parameter.
There exist also a continuum spectrum of basic solutions of which
we do not examine here as they are nonanalytic on the axis of
rotation. The stationary part of multipolar solutions that are
asymptotically flat we call external multipoles, and the
stationary part of multipolar solutions that are not
asymptotically flat we call internal multipoles. We showed that
external multipoles, and any asymptotically flat space-times of
\vsf, are massless no matter what is pasted smoothly as an
alternative internal completion of the space-times. Despite the
fact, some of the space-times can have non-vanishing total angular
momentum.

As in standard gravity a broad class of solutions can be
decomposed to the multipoles. Even though the resulting
space-times are solutions of Einstein equations of a
differentially rotating dust, thus physically more interesting
than van Stockum-Bonnor flows \cite{bib:vanstock},
\cite{bib:bonnor}, \cite{bib:our}, it seems that they are of no
physical concern as local energy density is not positive definite.
Anyway, we decided to publish the solutions as they present
another class of exactly solvable dust flows in general
relativity. Similarly, as the exactly solvable van Stockum--Bonnor
class of asymptotically flat flows along time Killing vector
field, the asymptotically flat external multipoles of \vsf{s},
which must have vanishing total mass irrespectively of internal
deformations\footnote{one can imagine such a deformation as a
replacement of the interior by a region of arbitrary space-time
such the resulting space-time be still smooth}, have no newtonian
limit and are useless in astrophysical situations. As far as
exotic forms of matter concern, the internal solutions of both the
models can be still considered as physically interesting and worth
to be examined if completed smoothly by, say, asymptotically flat
external (Kerr) vacuum solutions.

It turns out that the structure function $K=\vec{\xi}\vec{\eta}$
of the flow is the same as for the van-Stockum Bonnor flow
\cite{bib:our}, nonetheless, both the flows and geometry of the
resulting space-times are qualitatively different. An
asymptotically flat van Stockum--Bonnor flow is rigid,
non-expanding, and has positive energy density proportional to the
square of the vorticity scalar which does not vanish. An
asymptotically flat \vsf is differential, non-expanding, locally
non-rotating, and has negative definite energy density
proportional to the square of the non-vanishing shear tensor. Dust
of van Stockum--Bonnor does not rotate with respect to asymptotic
observers, it has a point-dependent physical velocity as measured
with respect to the locally dragged inertial frames and which is
proportional to $K$, whereas dust of a \vsf rotates differentially
with respect to asymptotic observers with angular velocity
proportional to $K$, and has vanishing physical velocity. The
peculiarities are possible due to purely relativistic effects and
show the flows have no newtonian limits.

As an example we constructed a $z$-symmetric, cylindrically
symmetric, and stationary for radii sufficiently large,
asymptotically flat space-time which is globally smooth (with the
exception of point-like singularities of curvature residing on the
axis of rotation), and which can be decomposed globally to the
external multipoles. The solution has nonzero total angular
momentum and zero total mass.

\section{Setup} We assume the global existence of the cylindrical
symmetry space-like Killing vector $\vec{\eta}$ with closed field
lines, together with the global existence of time translation
Killing vector $\vec{\xi}$ of which field lines are opened,
time-like for radii sufficiently large, and asymptotically
normalizable to unity. Having in mind stationary and
asymptotically flat space-times with cylindrical symmetry, we
shall proceed as in the standard theory of rotating stars in the
relativistic astrophysics \cite{bib:bardeen}. The assumption of
asymptotical flatness allows for the unique determination of
Killing vectors $\vec{\xi}$ and $\vec{\eta}$ by the defining
properties. In addition one assumes that asymptotically
$\vec{\xi}\vec{\eta}\to0$. Field lines of the Killing vectors,
which clearly are frame independent objects, may be viewed as two
of four coordinate lines in some particular coordinates in which
the time coordinate $t$ runs along open lines of $\vec{\xi}$ and
the cyclic coordinate $\phi$ along closed lines of $\vec{\eta}$,
that is, $\vec{\xi}\equiv\vec{\partial}_t$ and
$\vec{\eta}\equiv\vec{\partial}_{\phi}$ by definition. The other
two, denoted by $\rho$ and $z$, are arbitrary internal coordinates
in a two dimensional subspace orthogonal to $\vec{\xi}$ and
$\vec{\eta}$. In this coordinates the
most general line element of a cylindrically symmetric and
stationary space-time is fully determined by four structure
functions $\nu(\rho,z)$, $\psi(\rho,z)$, $\tilde{\omega}(\rho,z)$
and $\mu(\rho,z)$ and reads \cite{bib:bardeen}
\begin{equation}\ud{s}^2=\exp{2\lambda}\ud{t}^2-\exp{2\psi}
\br{\ud{\phi}-\tilde{\omega}\ud{t}}^2-\exp{2\mu}\br{\ud{
\rho}^2+\ud{z}^2}.\label{eq:carter_metric}
\end{equation}
 Bardeen coordinates are distinguished by the property
that in these coordinates Killing vectors $\vec{\xi}$ and
$\vec{\eta}$ attain particularly simple form
$$ \xi^{\mu}=\delta^{\mu}_t, \qquad \eta^{\mu}=\delta^{\mu}_{\phi}.$$
In asymptotically flat space-time one can introduce cylindrical
coordinate system in which the line element at infinity reduces to
$$\ud{s}^2\to\ud{t}^2-\ud{\rho}^2-\rho^2\ud{\phi}^2-\ud{z}^2$$
Axis of the coordinates frame are attached to 'fixed stars',
otherwise the axis would rotate. This in turn would be in
contradiction with the assumption that $\vec{\partial}_t$ can be
asymptotically normalized to unity. One assumes, therefore, that
asymptotically condition
$\exp{2\lambda}-\tilde{\omega}^2\exp{2\psi}>0$ should hold. The
latter is not the case e.g. in a rotating frame of reference that
rotates with constant angular velocity. Put differently, Bardeen
coordinates are asymptotically inertial (anyway, this metric can be treated more formally and be used to describe solutions that are not asymptotically flat).

As it turns out later, \vsf can be defined globally, that is,
locally non-rotating observers exist for all points (apart from
singularities), but, unlike for van-Stockum Bonnor flow,
space-time of \vsf is not globally stationary. The \vsf is
differential (the shear tensor does not vanish). The angular
velocity of asymptotically flat \vsf is, by construction,
identical to the angular velocity of dragging of inertial frames
$\om$. As so, dust rotates on circular orbits with respect to
asymptotical stationary observers even though it locally does not
rotate (the vorticity vector of the flow vanishes identically) and
its physical velocity, measured with respect to dragged local
inertial frames, is also identically zero.
\subsection{The line element of {\vsf}}
We derive the line element of the \vsf step by step from the
general form of the line element given by \ref{eq:carter_metric}.
We specify the structure functions uniquely by the requirement
that dust space-time trajectories are identical with field lines of
four-velocity field of locally non-rotating observers and that
Einstein equations are satisfied.
\subsubsection{A note on locally non-rotating observers}
It should be clear that irrespectively of any particular reference
frame, a generic cylindrically symmetric and stationary flow is
fully determined by a four-velocity field
$$\vec{U}(\Omega)=Z(\Omega)\br{\vec{\xi}+\Omega\vec{\eta}},$$
provided
$$ Z(\Omega)^{-2}\equiv\vec{\xi}\vec{\xi}
+2\Omega\vec{\xi}\vec{\eta}+\Omega^2\vec{\eta}\vec{\eta}>0,\quad
\pounds_{\vec{\xi}}\Omega=0=\pounds_{\vec{\eta}}\Omega.$$ In an
asymptotically flat space-time $\Omega$ has the interpretation of
the angular velocity measured with respect to 'fixed stars'. In particular,
the condition $\vec{U}(\tilde{\Omega})\vec{\eta}=0$ gives
$$\tilde{\Omega}=-\frac{\vec{\xi}\vec{\eta}}{\vec{\eta}\vec{\eta}}
\equiv\tilde{\omega}.$$ The vorticity tensor\footnote{see footnote
\ref{foot:def} } of velocity field
$\vec{n}\equiv{}\vec{U}(\tilde{\omega})$ vanishes identically,
which can be checked by a direct calculation in Bardeen
coordinates \ref{eq:carter_metric}. Therefore, the observers are
called locally non-rotating, although they move differentially on
circular orbits with respect to 'fixed stars' with angular
velocity $\tilde{\omega}$ of dragging of inertial frames. Indeed,
congruence of locally non-rotating observers distorts without
changing proper volume\footnote{Dilation tensor
$\vec{\Theta}(\vec{u})$ and (traceless) shear tensor
$\vec{\sigma}(\vec{u})$ are defined for a velocity field $\vec{u}$
as $\Theta_{\mu\nu}=\nabla_{\alpha}u^{\alpha}h_{\mu\nu}$ and
$\sigma_{\mu\nu}=\nabla_{(\alpha}u_{\beta)}h^{\alpha}_{
\phantom{\alpha}\mu}h^{\beta}_{
\phantom{\beta}\nu}-\frac{1}{3}\nabla_{\alpha}u^{\alpha}
h_{\mu\nu},$
 where $h^{\mu}_{\phantom{\mu}\nu}=
 \delta^{\mu}_{\phantom{\mu}\nu}-u^{\mu}u_{\nu}$ is projector $\vec{h}(\vec{u})$
 onto subspace orthogonal to $\vec{u}$. Vorticity tensor $\vec{\Omega}(\vec{u})$ for $\vec{u}$ is defined as
 $\omega_{\mu\nu}=\nabla_{[\alpha}u_{\beta]}h^{\alpha}_{
\phantom{\alpha}\mu}h^{\beta}_{ \phantom{\beta}\nu}$ and it yields
a derivative quantity that characterizes vertex -- the vorticity
vector
$\omega^{\mu}=\frac{1}{2}\frac{\epsilon^{\mu\alpha\beta\gamma}}{\sqrt{-g}}
u_{\alpha}\omega_{\beta\gamma}$. One defines also square of
dilation scalar
$\vec{\sigma}^2(\vec{u})=\frac{1}{2}\sigma^{\mu\nu}\sigma_{\mu\nu}\geq0$
and square of vorticity scalar
$\vec{\Omega}^2(\vec{u})=\frac{1}{2}\omega^{\mu\nu}\omega_{\mu\nu}\geq0$,
then
$\vec{\Omega}^2(\vec{u})=-\omega^{\mu}\omega_{\mu}$.\label{foot:def}}
$$\vec{\Omega}^2\br{\vec{n}}\equiv0,\qquad
\vec{\Theta}\br{\vec{n}}\equiv0,
\qquad\vec{\sigma}^2\br{\vec{n}}=-\frac{1}{4}\frac{\br{\vec{\eta}\vec{\eta}}^2
\br{\vec{\nabla}{\tilde{\omega}}}^2}{\left|\begin{array}{cc}
\vec{\xi}\vec{\eta}&\vec{\xi}\vec{\xi}\\
\vec{\eta}\vec{\eta}&\vec{\xi}\vec{\eta}\\
\end{array}\right|},\qquad \tilde{\omega}=
-\frac{\vec{\xi}\vec{\eta}}{\vec{\eta}\vec{\eta}}.$$  Local
inertial frames tangent to $\vec{n}$ play the role of local
standards of rest with respect to which 'physical' velocity is
measured. Therefore, the physical velocity of a particle moving
with four-velocity $\vec{U}(\Omega)$ is the relative velocity of
two inertial observers momentarily co-moving with
$\vec{U}(\Omega)$ and $\vec{n}$, respectively. After Special
Relativity we define the relative velocity as the hyperbolic
tangent of the hyperbolic angle $\chi$ between four-velocities
$\vec{U}(\Omega)$ and $\vec{n}$. In Bardeen coordinates it reads
\begin{equation}\tanh{\chi}=\exp{\psi-\lambda}(\Omega-\tilde{\omega})
.\label{eq:gen_vel}\end{equation} The result agrees with that of
\cite{bib:bardeen} where more 'physical' derivation was presented.
\subsubsection{Specifying the line element}
The energy-momentum tensor of dust matter moving along field lines
of vector field $\vec{n}$ reads
$\vec{T}=\mathcal{D}\vec{U}\otimes\vec{U}$ where
$$\vec{U}(\om)=Z(\om)\br{\vec{\xi}+\om\vec{\eta}}\qquad
\Rightarrow\quad U^{\mu}=e^{-\lambda} \sq{1,0,\om,0}.$$ From the
previous construction it follows that dust in asymptotically flat
\vsf moves differentially on circular orbits with angular velocity
of dragging of inertial frames $\om$ as seen by an asymptotic
stationary observer. This flow has vanishing vorticity vector,
that is, it does not rotate locally, and has no physical velocity
-- it is at rest with respect to congruence of local standards of
rest dragged with angular velocity $\om$. It follows also that
angular momentum per mass element of the flow $-U^{\mu}\eta_{\mu}$
is identically zero as $\vec{U}\equiv\vec{n}$ and $\vec{\eta}$ are
orthogonal. Nevertheless, as we shall show later, total angular
momentum does not vanish for particular kinds of asymptotically
flat space-times of \vsf which must contain singularities that contribute total angular momentum of the space-times.

Einstein's equations $\vec{G}=\kappa\vec{T}$ and the contracted
Bianchi identities $\nabla^{\mu}G_{\mu\nu}=0$ yield local
conservation law $\nabla^{\mu}T_{\mu\nu}=0$ that for dust gives
continuous flow along geodesic paths
$$\nabla_{\mu}\br{\mathcal{D}U^{\mu}}=0
\qquad \mathrm{and} \qquad U^{\nu}\nabla_{\nu}U_{\mu}=0.$$ The
continuity equation is satisfied identically for {\vsf} on the
power of Killing equations and of symmetries of energy density
$\vec{\xi}\mathcal{D}=0$ and $\vec{\eta}\mathcal{D}=0$. In Bardeen coordinates $
U^{\nu}\nabla_{\nu}U_{\mu}=\{0,-\partial_{\rho}\lambda,0,-\partial_z\lambda\}$
which implies that $\lambda$ must be a constant. Another
constraint on structure functions follows from the linear
combination of Einstein's equations
$$0\equiv\xi^{[\mu}\eta^{\alpha]}\xi_{[\alpha}\eta_{\nu]}
 \br{R_{\mu}^{\phantom{\mu}\nu}-\kappa\br{T_{\mu}^{\phantom{\mu}\nu}
 -\frac{1}{2}T\delta_{\mu}^{\phantom{\mu}\nu}}}
 =\frac{1}{4}\exp{\lambda+\psi-2\mu}\br{\partial_{\rho}^2+\partial_{z}^2}
\exp{\lambda+\psi}.$$ The left side is identically zero on the
power of Einstein equations, while the expression to the right
holds for pressureless perfect fluid in Bardeen coordinates. We
have not yet specified the arbitrary coordinates $\rho$ and $z$.
Asymptotical flatness in cylindrical coordinates requires that
$\exp{2\lambda}\to1$ and $\exp{2\psi}\to\rho^2$, thus
$\exp{\lambda+\psi}\to\rho$. The simplest way to concord the facts
is to choose as a solution of the above equation
\begin{equation}
\lambda=0,\qquad \exp{\psi}=\rho\label{eq:constr2}.\end{equation}
We have thus shown that {\vsf} requires at most two structure
functions $\tilde{\omega}$ and $\mu$.

Killing vector $\vec{\eta}$ is always space-like
$\vec{\eta}\vec{\eta}\equiv-\rho^2$ thus the space-time of \vsf is
globally cylindrically symmetric. As
$\vec{\xi}\vec{\xi}=1-\om^2\rho^2$ it is also stationary in
regions where $\om^2\rho^2<1$. Note that
$\det[g]=-\rho^2\exp{4\mu}$ does not depend explicitly on $\om$
thus the boundary surface $\om^2\rho^2=1$, which demarcates
stationary and non-stationary regions of a space-time, is
nonsingular, and in a sense, is alike to Schwarzschild horizon.

We stress the construction of the energy-momentum tensor of \vsf
makes sense globally as $U^{\mu}U_{\mu}>0$, or $\vec{\xi}\vec{\xi}
+2\om\vec{\xi}\vec{\eta}+\om^2\vec{\eta}\vec{\eta}\equiv1>0$ for
\vsf. $\vec{U}(\om)$ is time-like even though in some regions the
resulting space-time may be non-stationary, and so, $\vec{U}(\om)$
be a combination of two space-like vectors. It follows that in
space-times of \vsf the locally non-rotating observers exist
globally.

Similarily as in \cite{bib:our} we define $K=\vec{\xi}\vec{\eta}$
then in Bardeen coordinates $K=\rho^2\om$, and $(K,\mu)$ can be
used in place of $(\tilde{\omega},\mu)$ as structure functions of
{\vsf}. We have thus reduced the line element
\ref{eq:carter_metric} of \vsf to
\begin{equation}
\ud{s}^2=\br{1-\frac{K^2(\rho,z)}{\rho^2}}\ud{t}^2+2K(\rho,z)\ud{t}\ud{\phi}
-\rho^2\ud{\phi}^2 -\exp{2\mu(\rho,z)}
\br{\ud{\rho}^2+\ud{z}^2}\label{eq:vs_metric}.\end{equation} We
have seen, that in space-times given by the line element,
world-lines of locally non-rotating observers are geodesics.

At a given fixed point one can diagonalize the line element by a
local linear transformation (local boost) of the base forms, say,
$\ud{\phi}\to\ud{\phi}+\om\ud{t}$. At the fixed point the line
element reduces simply to
$\ud{t}^2-\rho^2\ud{\phi}^2-\exp{2\mu}\br{\ud{\rho}^2+\ud{z}^2}$.
The new base overlaps with the inertial frame of an observer
momentarily co-moving with the locally non-rotating observer at
that point.

\subsection{Equations for structure functions of {\vsf}}

We have shown that for {\vsf} general metric
\ref{eq:carter_metric} can be reduced to \ref{eq:vs_metric}. In
what follows we shall derive equations for $K$ and $\mu$. Let
$\vec{E}=\vec{G}-\kappa\vec{T}$ with
$\vec{T}=\mathcal{D}\vec{U}\otimes\vec{U}$, $\vec{U}=\vec{n}$.
Einstein's equation $\vec{E}=0$ imply from
$$\upd{E}{\rho}{\rho}=\frac{\exp{-2\mu}}{\rho}
\br{\rho^3\frac{\om_{,z}^2-\om_{,\rho}^2}{4}-\mu_{,\rho}},\qquad
\upd{E}{\rho}{z}=-
\frac{\exp{-2\mu}}{\rho}\br{\rho^3\frac{\om_{,\rho}\om_{,z}}{2}+\mu_{,z}}$$
($\upd{E}{\rho}{\rho}=-\upd{E}{z}{z}$,
$\upd{E}{\rho}{z}=\upd{E}{z}{\rho}$)
 that
\begin{equation}\mu_{,\rho}=\rho^3\frac{\om_{,z}^2-\om_{,\rho}^2}{4}, \qquad
\mu_{,z} =-\rho^3\frac{\om_{,\rho}\om_{,z}}{2}.
\label{eq:nu}\end{equation} For $\mathcal{C}^2$ solutions the
Schwarz identity $\mu_{,\rho{z}}=\mu_{,z\rho}$ imposes on $\om$
the linear elliptic constraint
\begin{equation}\om_{,\rho\rho}+3\rho^{-1}\om_{,\rho}+\om_{,zz}=0
\quad\Leftrightarrow\quad
K_{,\rho\rho}-\frac{K_{,\rho}}{\rho}+K_{,zz}=0.
\label{eq:n}\end{equation} As
$$\upd{E}{t}{\phi}=\frac{\exp{-2\mu}}{2}
\br{K_{,\rho\rho}-\frac{K_{,\rho}}{\rho}+K_{,zz}}$$ equation
$\upd{E}{t}{\phi}=0$ is satisfied  identically.
 By
calculating $\mu_{,\rho\rho}$ and $\mu_{,zz}$ from \ref{eq:nu} and
using \ref{eq:n}, one may check the component $\upd{E}{t}{t}$
(then $\upd{E}{\phi}{t}=K\rho^{-2}\upd{E}{t}{t}$) reduces to
$\upd{E}{t}{t}=-\exp{-2\mu}\rho^2
\br{\om_{,\rho}^2+\om_{,z}^2}-\kappa\mathcal{D}=0$,
and finally,
\begin{equation}\kappa\mathcal{D}=-\exp{-2\mu}\rho^2
\br{\om_{,\rho}^2+\om_{,z}^2}.\label{eq:dens}
\end{equation}The other components of $\vec{E}$ vanish identically by symmetry. As energy density is negative definite \vsf is unphysical in the sense it cannot be made of ordinary matter.
Once a solution of \ref{eq:n} is found, which is simple task due
to linearity, \ref{eq:dens} gives the respective energy density
and \ref{eq:nu} can be easily integrated. Contribution to the
total mass, from regions where $\om\in\mathcal{C}^1$, is negative
and given for solutions by
\begin{equation}M_{smooth}=-\frac{c^2}{4G}\int
\rho^3\br{\om_{,\rho}^2+\om_{,z}^2}\ud{\rho}\ud{z}\label{eq:msmooth}\end{equation}
anyway, we shall show that asymptotically flat space-times of \vsf
must contain distributional sources of a nett positive mass
$-M_{smooth}$ that cancels contributions (possibly infinite) from
smooth components given by integrals as the above, as the
asymptotic total mass, calculated as a surface integral at
infinity, vanishes for asymptotically flat \vsf{s}.

Note, that elliptic equation \ref{eq:n} is the condition for
critical points $\om$ of energy functional \ref{eq:msmooth} (when
derived by integrating \ref{eq:dens} with respect to the proper
volume $\exp{2\mu}\rho$ of the surface of constant time, then
$M_{smooth}=\int\mathcal{D}\exp{2\mu}\rho
\ud{\rho}\ud{\phi}\ud{z}$) in any finite region of a hyper-surface
of constant time bounded by a closed line in the plane $(\rho,z)$
and such that $\om$ is at least $\mathcal{C}^{1}$ (piecewise) in
the region, namely
$$\delta M[\om]=0\qquad \Rightarrow\qquad \om_{,\rho\rho}+3\frac{\om_{,\rho}}{\rho}+\om_{,zz}=0.$$ For the solutions the maximum
(minimum) principle applies.
\subsection{Some solutions}
To obtain basic solutions we transform \ref{eq:n} to spherical
coordinates $\rho\to{}r\sin{\theta}$, $z\to{}r\cos{\theta}$. By
substituting $K(r,\theta)=R(r)Y(\theta)$, the separation of
variables gives $R(r)=r^{-n}$ or $R(r)=r^{n+1}$, and the
hypergeometric equation for $Y(x)$, where $x=\cos{\theta}$. We
assume here $n\in\mathbb{N}$ and take only solutions that are
analytic at $x=\pm1$, by which the multipole series is
established. The general formula for $z$-antisymmetric and
$z$-symmetric external multipoles that are solutions of \ref{eq:n}
and that give asymptotically flat space-times is
\begin{equation}K(\rho,z)=\left\{\begin{array}{cc}
\frac{1}{\br{\rho^2+z^2}^{m-1/2}} \
_2F_1\br{m-\frac{1}{2},-m;\frac{1}{2},
\frac{z^2}{\rho^2+z^2}},&m=1,2,3,\dots\\
\frac{z}{\br{\rho^2+z^2}^{m+1/2}} \
_2F_1\br{m+\frac{1}{2},-m;\frac{3}{2},
\frac{z^2}{\rho^2+z^2}},&m=1,2,3,\dots
\end{array}\right.\label{eq:monopoles}\end{equation}
For $m=0$ one obtains $K=\sqrt{\rho^2+z^2}$ and the
monopole\footnote{We assume the same nomenclature for the $K$ as
in \cite{bib:our} where Euclidean mass density
$\rho^{-2}\br{K_{,\rho}^2+K_{,z}^2}$ for the monopole of the van
Stockum-Bonnor flow was spherically symmetric, even though, as one
can check, the Euclidean mass density
$-\rho^{+2}\br{\om_{,\rho}^2+\om_{,z}^2}$ for the monopole
$\om=K/\rho^2$ of the \vsf is not spherically symmetric.}
$K=z/\sqrt{\rho^2+z^2}$ which are non-flat asymptotically . The
corresponding series of internal multipoles (singular at infinity)
is \begin{equation}K(\rho,z)=\left\{\begin{array}{cc}
\br{\rho^2+z^2}^{m} \ _2F_1\br{m-\frac{1}{2},-m;\frac{1}{2},
\frac{z^2}{\rho^2+z^2}},&m=0,1,2,\dots\\
z\br{\rho^2+z^2}^{m} \ _2F_1\br{m+\frac{1}{2},-m;\frac{3}{2},
\frac{z^2}{\rho^2+z^2}},&m=0,1,2,\dots
\end{array}\right.\label{eq:monopo2}\end{equation}
\subsubsection{An example}
In the previous paper we found an example of asymptotically flat
solution (note that equation for $K$ function are identical but the models and line elements are essentially different)
\begin{eqnarray}K_{a}(\rho,z)=\lefteqn{\frac{2\rho^2z}{
\br{z+a}\sqrt{\rho^2+\br{z-a}^2}+\br{z-a}\sqrt{\rho^2+\br{z+a}^2}
}+\dots{}}\nonumber \\ & & {}\dots+\frac{\rho^2}{2a}
\ln\br{\frac{z-a+\sqrt{\rho^2+\br{z-a}^2}}{z+a+\sqrt{
\rho^2+\br{z+a}^2}}}{} \label{eq:my_sol}\end{eqnarray} which can
be expanded in the series of external $z$-symmetric multipoles
$$K_{a}(\rho,z)=\frac{2}{3}\frac{\rho^2}{r^3}a^2-\frac{1}{5}
\frac{\rho^2\br{\rho^2-4z^2}}{r^7}a^4+\frac{3}{28}\frac{\rho^2
\br{\rho^4-12\rho^2z^2+8z^4}}{r^{11}}a^6+\dots,$$ where
$r=\sqrt{\rho^2+z^2}$. Solution $K_a$ has single extremum
$K(0,0)=a$ and, apart from the singular points $(0,a)$ and
$(0,-a)$, is everywhere smooth. The two points are singular since,
for example, $\partial_zK$ as a function of $z$, for $\rho=0$ has
a jump at $z=\pm{}a$, $a>0$. The corresponding angular velocity
$\om_a=\rho^{-2}K_a$ has additional singularities, namely it is
analytical everywhere with the exception of the closed segment
$\rho=0$, $z\in[-a,a]$. The resulting space-time is asymptotically
flat
$$K_a(r,\theta)\sim\frac{2}{3}\frac{a^2}{r}\sin^2{\theta},\quad
\om_a(r,\theta)\sim\frac{2}{3}\frac{a^2}{r^3} \qquad
\mu_a(r,\theta)\sim\frac{a^4\sin^4{\theta}}{4r^4},\qquad
r\to\infty,$$ and
\begin{eqnarray*}
&g_{tt}\sim1-\frac{4}{9}\frac{a^4}{r^4}\sin^2{\theta},\quad
g_{t\phi}\sim\frac{2}{3}\frac{a^2}{r}\sin^2{\theta}\br{1+\frac{3}{20}
\frac{a^2}{r^2}\br{3+5\cos{2\theta}}},&\\
&g_{\phi\phi}=-r^2\sin^2{\theta},\quad
g_{rr}=\frac{g_{\theta\theta}}{r^2}
\sim-1-\frac{a^4}{2r^4}\sin^4{\theta}.&\end{eqnarray*}Comparison
with asymptotical expansion of the Kerr metric allows to determine
total mass of the space-time, which is $M=0$, and total angular
momentum, which is $J=a^2/3$. The result proves that the singular
segment of solution $\om_a$, located on the $z$-axis at
$-a\le{}z\le{}a$, are sources of positive mass that cancel
negative mass of the remaining regions ($\mathcal{D}<0$). In
general, total mass of asymptotically flat space-time can be
calculated as surface integral over sphere $\mathcal{S}^2$ at
infinity, and for the line element \ref{eq:vs_metric}
\begin{equation}M=\lim_{r\to\infty}\frac{1}{8\pi}
\int\limits_{\mathcal{S}^2}\star\ud{\vec{\xi}}=-\lim_{r\to\infty}\frac{1}{8\pi}
\iint\br{r^2\sin^2{\theta}\om\partial_{r}\om}
r^2\sin{\theta}\ud{\theta}\wedge\ud{\phi},\label{eq:gauss_m}\end{equation}
where $\vec{\xi}=\xi_{\mu}\ud{x}^{\mu}$ and $\star$ is the Hodge
operator. For $r$ finite and sufficiently large the mass function
is positive and behaves as $M(r)\sim4a^4/\br{9r^3}>0$. The
function tends to $0$ as $r\to\infty$, which confirms our
statement. The singular segment is the space-time singularity as
curvature scalar is proportional to $\mathcal{D}$ which does not
exist as a function on the segment, or more precisely,
$\mathcal{D}$ is a distribution. Total angular momentum of an
asymptotically flat space-time can be calculated as a surface
integral over the sphere $\mathcal{S}^2$ at infinity, and for the
line element \ref{eq:vs_metric}
\begin{equation}J=-\lim_{r\to\infty}\frac{1}{16\pi}
\int\limits_{\mathcal{S}^2}\star\ud{\vec{\eta}}=
-\lim_{r\to\infty}\frac{1}{16\pi}\iint
\br{r^2\sin^2{\theta}\partial_r\om}r^2\sin{\theta}
\ud{\theta}\wedge\ud{\phi}\label{eq:gauss_j},\end{equation} and we
again obtain $J=a^2/3$ for solution \ref{eq:my_sol}. As we have
mentioned earlier, the specific angular momentum for any regular
element of the flow vanishes identically as trajectories of the
flow are world-lines of locally non-rotating observers. Another
way of seeing this is to calculate the form
$\ud\star\ud{\vec{\eta}}$, which measures angular momentum density
when one take $\ud{t}=0$, and which as a general form vanishes
identically for $\om$ satisfying equation \ref{eq:n}. In the
singular regions where $\om$ is a distribution and does not
satisfy \ref{eq:n}, the contribution from
$\ud\star\ud{\vec{\eta}}$ to total angular momentum can be
calculated from the above surface integral.

\subsection{Asymptotically flat space-times of {\vsf} are massless,
some of them may have non-vanishing total angular momentum} The
effect of vanishing of total mass should hold for all
asymptotically flat space-times of \vsf. Asymptotical flatness for
line element \ref{eq:vs_metric} means that $\om$ falls off like
$r^{-3}$ or quicker in spherical coordinates. If $\om\sim{}r^{-3}$
then such space-time may have nonzero total angular momentum.
Vanishing of total mass is due to the presence of singularities
that contribute positive mass that cancel contribution from
regular regions of the space-times where energy density is
negative definite. Similarly, the existence of asymptotically flat
\vsf{s} implies that the singularities they contain must have
finite angular momentum equal to the total angular momentum of the
space-times as the flow contributes no angular momentum in the
regular regions.

\section{Conclusions}
We examined cylindrically symmetric and stationary dust flow along
world-lines of locally non-rotating observers of which space
trajectories are rings about common axis of rotation. Such flow
can be constructed globally with the exception of regions where
curvature singularities reside and that are additional sources of
gravitational field. The flow is differential (the shear tensor is
non-zero), non-expanding (the dilation scalar vanishes
identically) and locally does not rotate (the vorticity tensor
vanishes identically). Local mass density of the flow is
necessarily negative-definite which makes the flow nonphysical as
far as usual forms of matter are concerned.

There exist asymptotically flat space-times of the flow.
Components of the flow move differentially on circular orbits with
respect to an asymptotic stationary observer. The space-times have
vanishing total mass and contain internal singularities where
distributional sources of positive mass and of angular momentum
are located, and this should be understood in the following sense.
It turns out that a surface integral over the sphere at infinity
and which reproduces total mass (of the Kerr black hole, for
example), vanishes for the solutions. For radii sufficiently large
the resulting mass function is positive and attains zero
monotonically in the limit of infinite radius. As local energy
density is negative definite in the regions where space-time is
smooth, the internal singularities of curvature the solutions
contain, must be distributional sources of positive mass (maybe
infinite) of which contribution to the total mass is screened by
the regular regions, such the mass function is zero at infinity.
This phenomenon is quite analogous with the screening of
singularities of negative mass by regions of positive energy
density of asymptotically flat van Stockum-Bonnor flow. By
construction of \vsf the specific angular momentum per particle is
zero, nevertheless, total angular momentum of a class of the
asymptotically flat space-times may be non-zero and we gave an
example in \ref{eq:my_sol}. As the \vsf has no specific angular
momentum, the total angular momentum, which is given by a surface
integral over the sphere at infinity and which reproduces total
angular momentum (of the Kerr black hole, for example), must be
located in singularities of the space-times.

Among other solutions, the model contains an infinite sequence of
smooth asymptotically flat multipolar solutions to which a class of other
asymptotically flat solutions can be decomposed. There also exist an
infinite sequence of the corresponding internal multipolar
solutions that are not asymptotically flat. The solutions are
given in general by formulas \ref{eq:monopoles} and
\ref{eq:monopo2}.

\section{Acknowledgements}

This research is partially supported by the Polish Ministry of Science and Informatization, grants no. 1 P03D 005 28 and no. 1 P03B 012 29.

\end{document}